\documentclass[namedreferences]{kluwer}

\newdisplay{guess}{Conjecture}

\usepackage[dvips]{graphicx}

\newcommand{\RSun}{R_\odot}
\newcommand{\MSun}{R_\odot}
\newcommand{\LSun}{R_\odot}
\newcommand{\kms}{\mathrm{km}\,\mathrm{s}^{-1}}
\newcommand{\logg}{\log \, (g/g_0)}

\begin{document}                                                                                   
\begin{article}

\begin{opening}         

\title{Disentangling effective temperatures of individual eclipsing binary components by means of color-index constraining}
\author{A.~\surname{Pr\v sa}}
\author{T.~\surname{Zwitter}}
\runningauthor{Pr\v sa, Zwitter}
\runningtitle{Disentangling individual temperatures of eclipsing binaries}
\institute{University of Ljubljana, Dept. of Physics, Jadranska 19, SI-1000 Ljubljana, EU}

\begin{abstract}
Eclipsing binary stars are gratifying objects because of their unique geometrical properties upon which all important physical parameters such as masses, radii, temperatures, luminosities and distance may be obtained in absolute scale. This poses strict demand on the model to be free of systematic effects that would influence the results later used for calibrations, catalogs and evolution theory. We present an objective scheme of obtaining individual temperatures of both binary system components by means of color-index constraining, with the only requirement that the observational data-set is acquired in a standard photometric system. We show that for a modest case of two similar main-sequence components the erroneous approach of \emph{assuming} the temperature of the primary star from the color index yields temperatures which are systematically wrong by $\sim 100\,$K.
\end{abstract}


\end{opening}      

\bigskip
\centerline{\bf 1. Introduction}
\bigskip

Eclipsing binary stars (EBs) are renowned for their typical geometrical layout and well-understood underlying physics. Solving the inverse problem for detached EBs has become an indispensable tool for obtaining absolute dimensions of individual stars (their masses, radii, temperatures, and luminosities). The impact is very broad: EBs are used to establish color calibrations \cite{harmanec1988,flower1996,popper1998}, evolutionary properties based on exact coevality of both components \cite{ribas2000,lastennet2002} and relations between $M$-$L$-$R$-$T$ quantities for different spectral types and luminosity classes \cite{gorda1998}. It is thus very important to model EBs as objectively as possible, otherwise it is likely that systematic errors may creep in and mislead any theoretical models built on those results. In this paper we concentrate on a particular issue of objectively determining effective temperatures of individual components in an EB system.

The common practice of determining effective temperatures of individual components is to \emph{assume} the temperature of one star and obtain the temperature of the other star by the model. The assumed temperature is usually obtained from observed color-indices at quarter-phase by using a color calibration or by spectral analysis and comparison against synthetic spectra database. While this approach may be adequate for components having significantly different luminosities, (so that the luminosity of one star is equal to the system luminosity for all practical purposes), it is very inadequate for components with similar luminosities and introduces systematical effects into model solutions. When dealing with photometric observations that are acquired in a standard photometric system (i.e.~Johnson, Cousins, Str\"omgren, \dots), effective temperatures of both components may be disentangled from two or more photometric light curves \emph{objectively}, without any assumptions. We build this argument and demonstrate its implications on modeling in the following Sections.

\bigskip
\centerline{\bf 2. Simulation}
\bigskip

To evaluate the uncertainty of determining a primary star temperature a priori, we built a synthetic binary star model. Testing the method against a synthetic model may seem artificial, but the obvious advantage of knowing the right solution is the only true way of both qualitative and quantitative assessment.

\begin{table}[t]
\caption{Physical parameters of the F8~V--G1~V test binary star. Linear ($x$) and non-linear ($y$) coefficients of the logarithmic limb darkening law for Johnson B and V passbands, taken from van Hamme (1993).}
\label{table_of_parameters}
\begin{tabular*}{\maxfloatwidth}{ll}
\includegraphics[width=4.5cm]{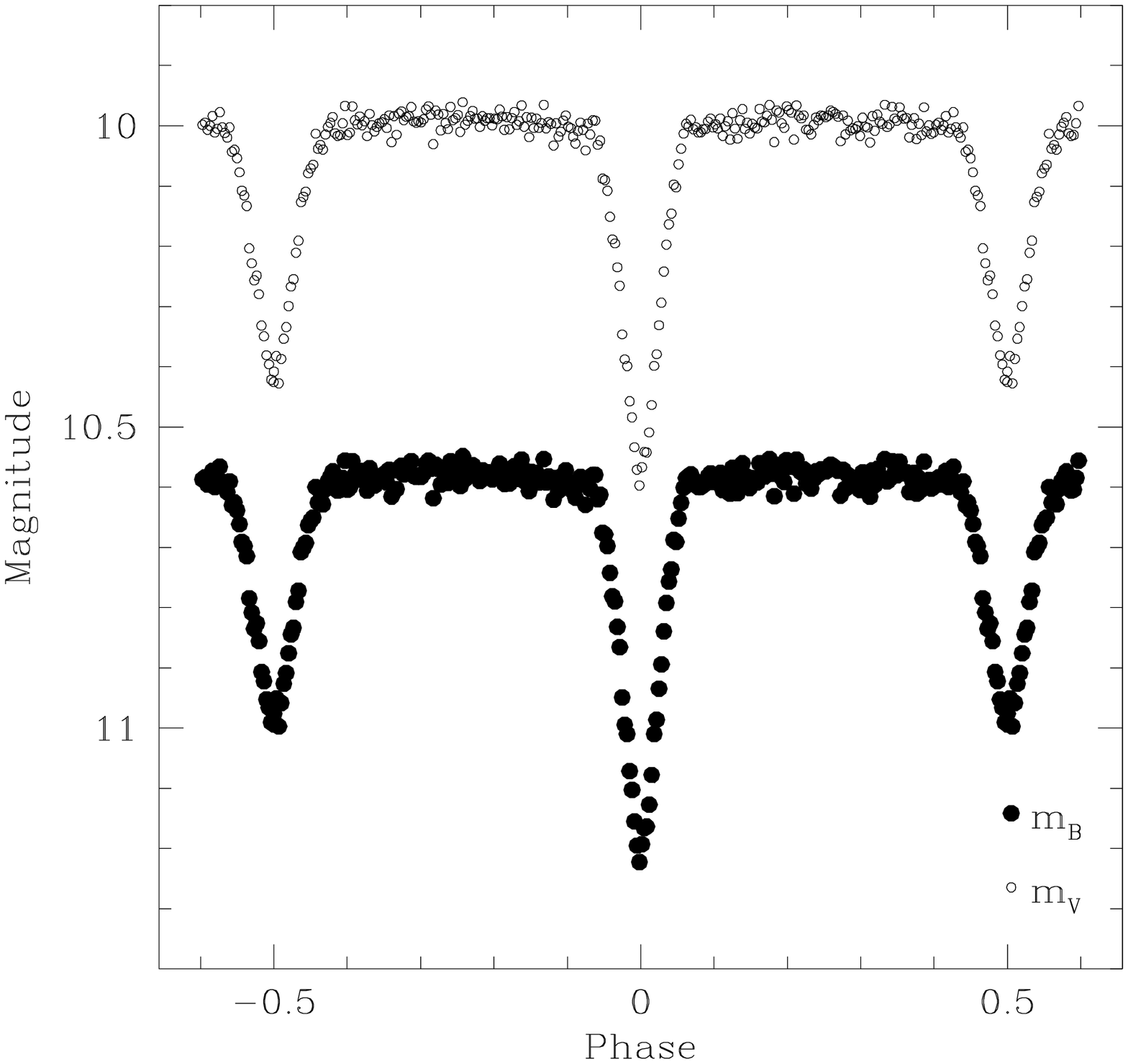} & \\
\includegraphics[width=4.5cm,height=2.25cm]{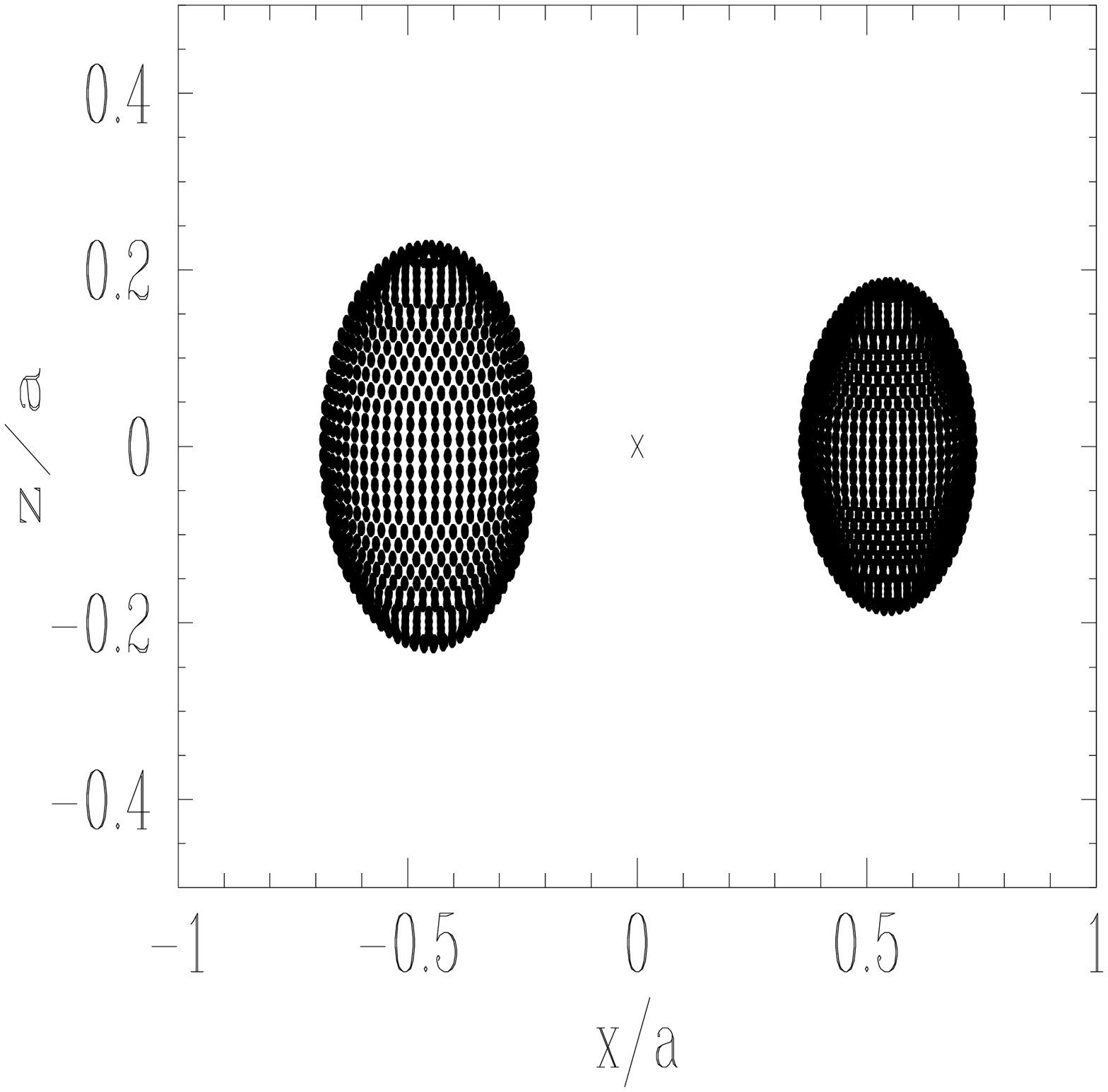} &
\raisebox{3.3cm}[0pt]{
\begin{tabular}{lccc}
\hline
Parameter [units]                 &	   & Binary &	     \\
                                  &  F8~V  &	    &  G1~V  \\
\hline
$P_0$ [days]                      &	   &  1.000 &	     \\
$a\,\,\,[\RSun]$                  &	   &  5.524 &	     \\
$q=m_2/m_1$                       &	   &  0.831 &	     \\
$i\,\,\,[{}^\circ]$               &	   & 85.000 &	     \\
$v_\gamma\,\,\,[\kms]$            &	   & 15.000 &	     \\
$T_\mathrm{eff}\,\,\,[\mathrm K]$ &   6200 &	    &	5860 \\
$L\,\,\,[\LSun]$                  &  2.100 &	    &  1.100 \\
$M\,\,\,[\MSun]$                  &  1.239 &	    &  1.030 \\
$R\,\,\,[\RSun]$                  &  1.260 &	    &  1.020 \\
$\Omega\,\,\,[-]^a$               &  5.244 &	    &  5.599 \\
$\logg\,\,\,[-]^b$                &  4.33  &	    &  4.43  \\
$x_B\,\,\,[-]$                    &  0.818 &	    &  0.833 \\
$y_B\,\,\,[-]$                    &  0.203 &	    &  0.158 \\
$x_V\,\,\,[-]$                    &  0.730 &	    &  0.753 \\
$y_V\,\,\,[-]$                    &  0.264 &	    &  0.242 \\
\hline
\multicolumn{4}{l}{${}^a$$\Omega$ is potential as defined in \inlinecite{wilson1979}.} \\
\multicolumn{4}{l}{${}^b$$g_0 = 1\,\mathrm{cm}\,\mathrm s^{-2}$ makes $g/g_0$ dimensionless.} \\
\end{tabular}
} \\
\end{tabular*}
\end{table}

%\nocite{vanhamme1993}

Our synthetic binary consists of two main-sequence F8~V--G1~V components with their most important orbital and physical parameters listed in Table \ref{table_of_parameters}. It is a partially eclipsing detached binary with only slight shape distortion of both components ($R_{1,\mathrm{pole}}/R_{1,\mathrm{point}} = 0.974$, $R_{2,\mathrm{pole}}/R_{2,\mathrm{point}} = 0.979$). Light curves are generated with {\tt PHOEBE} \cite{prsa2005} for Johnson B and V passbands in 300 phase points with Poissonian scatter $\sigma_{\mathrm{LC}} = 0.015$ and quarter-phase magnitude $m_V = 10.0$. Passband transmission curves were taken from ADPS \cite{moro2000}.

%{\tt PHOEBE} inherits its modeling principles from the \inlinecite{wilson1971} code {\tt WD}. %Before we get into details about the actual simulation, we should point out a common point of confusion when working with {\tt PHOEBE} and {\tt WD}: how is the total flux that is computed by the model normalized and how is it connected to passband luminosity $L_1^i$? The flux at reference phase (usually quarter-phase), in absence of 3rd light, is given by:
%
%\begin{equation} \label{total_flux}
%J_{\mathrm{ref}}^i = \frac{L_1^i}{4\pi} \left( 1 + \frac{L_2^i}{L_1^i} \right),
%\end{equation}
%
%where $L_1^i$ and $L_2^i$ are passband luminosities of the primary and secondary component in the $i$-th passband, respectively, and the term $4\pi$ comes from the definition of emergent intensity per steradian's worth of area. This relative flux can be expressed in physical units only after the distance to the binary and at least one of the temperatures are known.

The simulation logic is as follows: we take Kurucz's spectral energy distribution (SED) function for both components from the spectra database compiled by \inlinecite{zwitter2004}, doppler-shift them to the reference phase, sum their SEDs weighted by their corresponding luminosity and convolve the sum with the passband response functions of Johnson B and V filters. This enables us to compute the B-V color-index, which needs to be zero-corrected with respect to spectral type A0 (adopted temperature $9560\,\mathrm K$). Once this is done, we have obtained a synthetic prediction of the color index. Alternatively, the same result may be obtained by using color calibrations, e.g.~\inlinecite{flower1996}'s corrected tables given by \inlinecite{prsa2005}. Color indices may now be readily transformed to the ratio of passband fluxes $J_\mathrm{ref}^V/J_\mathrm{ref}^B$. 

Once the light curves are built so that the color index is preserved, we follow the color-index constraining principle described by \inlinecite{prsa2005} to assess the implications of erroneous \emph{assumed} value of $T_\mathrm{eff1}$.  
\bigskip
\centerline{\bf 3. Results}
\bigskip

Our analysis brings us to the following conclusions:

\begin{enumerate}
\item Individual temperatures without the imposed color-index con\-stra\-int are fully correlated, as is well known. The reason for this is the degeneracy between the temperatures and passband luminosities $L_1^i$: local effects of slightly different temperatures on the \emph{shape} of the light curve are very small, hardly noticeable (they scale roughly with the ratio of both temperatures). However, even slightly different temperatures cause an additional effect which is an order of magnitude larger: they scale both light curves, thus changing the color index. If this change is not constrained, the model relies only on secular effects, which are almost always buried in data noise and parameter degeneracy. Without color-constraining there is thus \emph{no practical way} to pinpoint a location in the $T_\mathrm{eff1}$-$T_\mathrm{eff2}$ cross-section that corresponds to the right solution.

\begin{figure}[t]
\begin{center}
\includegraphics[width=12cm,height=6cm]{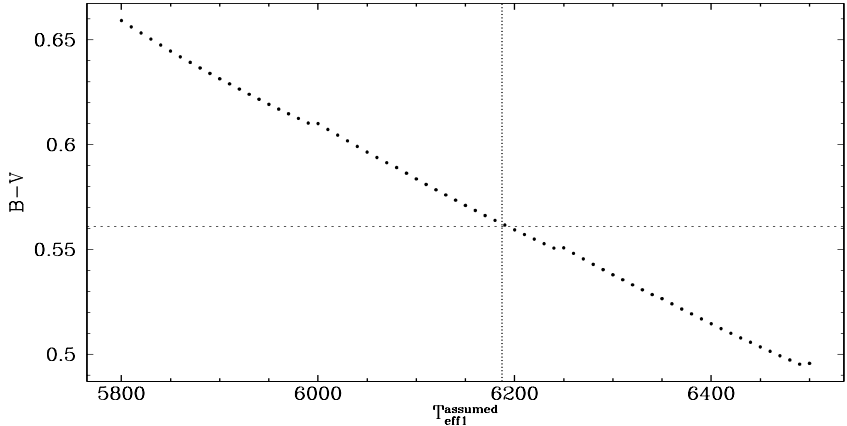} \\
\end{center}
\caption{Color index dependence on the assumed value of the primary star temperature. If color-constraining is not applied, all depicted points may be satisfied by the model. If we are to assume the value of $T_\mathrm{eff1}$ from the color index at quarter-phase, it would yield $T_\mathrm{eff1} = 6080\,$K, a value which is $\sim$120\,K off from the correct value. If on the other hand we apply color-constraining, we immediately get a handle on which color index corresponds to the temperatures yielded by the model. Cross-hairs denote the solution obtained by color-constraining, which gives $T_\mathrm{eff1} = 6187 \pm 32\,$K.}
\label{bv_teff_assumed}
\end{figure}

\item Unless the luminosity of one component is overwhelming, the \emph{assumed} value of the temperature is doubtful. Fig.~\ref{bv_teff_assumed} shows a range of assumed values for $T_\mathrm{eff1}$ and the color indices yielded by the model that is \emph{not} color-constrained. Since shapes of the light curves determine the ratio of luminosities of individual stars, it has to remain approximately constant. To achieve this with changing temperatures, the model adapts passband luminosities (first order effect) and stellar radii (second order effect) accordingly. The problem now becomes immediately evident: if we are to assume a primary star temperature from the quarter-phase color index, it would be roughly $T_\mathrm{eff1}^\mathrm{assumed} = 6080\,\mathrm K$, around $120\,$K off just because the luminosity of the secondary star is not negligible. In principle, the solution could then be \emph{manually} re-iterated by computing the predicted color-index from the solution, comparing it to the observed color index and then modifying the assumed temperature to get closer to the right solution. This is tedious and prone to subjective considerations, it should thus be avoided.

\item Adopting the color-constraint means keeping the ratio of passband luminosities $L_1^V/L_1^B$ accordant with the observed color index. This means that \emph{only} $L_1^V$ is adjusted to reproduce the $V$ light curve and $L_1^B$ (and other $L_1^i$ for the remaining passbands) is then computed from this constraint. The degeneracy between the temperatures and the passband luminosities is now broken: we obtain a straight horizontal line over Fig.~\ref{bv_teff_assumed} which best represents the color-index. Thus, we obtain the primary temperature \emph{without any assumptions}, and the secondary temperature follows directly from the model. Applying this to our test binary, we obtain individual temperatures $T_\mathrm{eff1} = 6\,187\,\,\mathrm K \pm 32$\,K and $T_\mathrm{eff2} = 5\,868\,\mathrm K \pm 31$\,K follows from the model. Comparing these values to true values $T_1 = 6\,200$\,K and $T_2 = 5\,860$\,K is very encouraging.

\begin{figure}[t]
\begin{center}
\includegraphics[width=10cm]{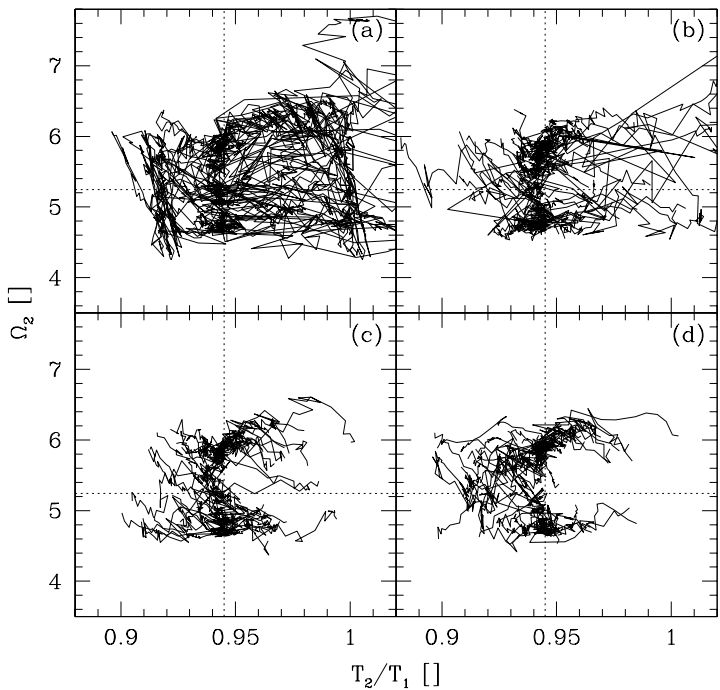} \\
\end{center}
\caption{Convergence tracer for the $T_\mathrm{eff2}/T_\mathrm{eff1}$-$\Omega_2$ cross-section. Cross-hairs denote the correct solution. Panels $(a)$ to $(d)$ correspond to the solution obtained by Nelder \& Mead's downhill Simplex (NMS) for 0, 1, 2 and 3 parameter kicks. See Pr\v sa \& Zwitter (2005) for details on interpreting convergence tracers.}
\label{tracer}
\end{figure}

\item Finally, let us consider the correlation between the temperatures and the radii of the binary system components. We have mentioned earlier that this is a second order effect, but it is still very important to quantify the correlation between those two parameters. Since the radii come into the model via surface potentials $\Omega$ (defined in \opencite{wilson1979}), we need to consider the correlation between the ratio of temperatures $T_2/T_1$ and surface potentials. Fig.~\ref{tracer} shows the convergence tracer for the $T_2/T_1$-$\Omega_2$ cross-section: tracing parameter values from each starting point, iteration after iteration, all the way to the converged solution. Attractors -- regions that attract most convergence traces -- within these cross-sections reveal parameter correlations and degeneracy. Panels denoted with $(a)$ to $(d)$ represent three subsequent parameter kicks, which are introduced and thoroughly explained by \inlinecite{prsa2005}.
\end{enumerate}

\bigskip
\centerline{\bf 4. Discussion}
\bigskip

We have demonstrated how severe the implications of assuming a primary star temperature from the color index may be if extra care is not taken to submit the acquired solution to manual re-iteration. The effect is already very pronounced for a modest case of two similar main-sequence components and it would be significantly augmented e.g.~for a hot main-sequence--cool giant system, where both stars have similar luminosities; in those cases the published solution may be off by as much as $\sim$1000\,K. By having the color-constraining method in {\tt PHOEBE}, we hope to raise awareness of this issue in future solution-seeking analyses.

\bibliography{mnemonic,colors}

\end{article}
\end{document}